# Time in the cell: a plausible role for the plasma membrane

Sepehr Ehsani
Department of Laboratory Medicine and Pathobiology, and Tanz Centre for Research in Neurodegenerative Diseases, University of Toronto, Toronto, Ontario, M5S 3H2, Canada
sepehr.ehsani@utoronto.ca

29 September 2012

**ABSTRACT**
*All cells must keep time to consistently perform vital biological functions. To that end, the coupling and interrelatedness of diverse subsecond events in the complex cellular environment, such as protein folding or translation rates, cannot simply result from the chance convergence of the inherent chemical properties of these phenomena, but may instead be synchronized through a cell-wide pacemaking mechanism. Picosecond vibrations of lipid membranes may play a role in such a mechanism.*

**MAIN TEXT**
Periodicity based on the twenty-four-hour daily cycle is essential to the normal functioning of many biological systems. Beginning with detailed genetic studies on fibroblast cells in the late 1990s (Balsalobre et al., 1998), research on circadian rhythms has flourished and circadian clocks have been shown to affect the expression of a considerable percentage of genes in the mammalian genome (Hughes et al., 2009). Some periodic events, such as limb development, do not need a clock (Mackem and Lewandoski, 2011), and in some cases of gene expression in the *Drosophila* embryo, the observed periodicity may be due to (at times indistinct) 'nonperiodic' processes (Roth and Panfilio, 2012). However, what is time on the cellular level, as distinct from time on the organism- or tissue-level? In other words, how does a cell keep time during the plethora of tasks it completes every second or minute? Furthermore, can synchrony amongst these tasks merely be due to the convergence of their reaction frequencies during evolution? Although single-cell studies have demonstrated variability and stochasticity from one cell to another (Li and Xie, 2011) and attempts at modeling noise between individual cells have progressed (Munsky et al., 2012), all healthy eukaryotic and prokaryotic cells seem to accomplish vital biological tasks with remarkable accuracy, precision and consistency. It is therefore not farfetched to assume that a fundamental unit of time exists across all cells to allow for a unified pace of biological functions. Where can the search for such a unit begin?

Cellular life began about 3.5 billion years ago, followed by the Great Oxidation Event a billion years later. Primitive cells at that time, for which counterparts in certain phyla still exist, must have followed orderly-paced pathways to survive. Given that essential biological tasks, such as transcription, translation and signalling pathways are, thematically, extremely conserved, any cellular-level pacemaker must have been simple enough to allow for its perpetuation across different phyla. Attempts hinting at this simplicity in biological order can be traced back to Erwin Schrödinger's proposition of an organism feeding upon negative entropy (Schrödinger, 1945), which can be complemented by the more contemporary realization that more complex biological modeling can at times lead to simpler outcomes (Szostak, 2011). Such a pacemaker can either be external or internal to the cell. In either case, it should fulfill two requirements: (i) possess a biologically-inherent oscillatory chemistry and (ii) be able to synchronize between and amongst cells of the same or different species.

A number of external forces can be tabulated as affecting a cell and providing contextual cues from the environment (**Figure 1**): temperature (Vanin et al., 2012), gravity/magnetism (Maeda et al., 2012), light/dark cycles, cell-cell contact forces (Serra-Picamal et al., 2012) and sound can be proposed. Due to the multitude of life environments, however, none of these external cues can provide a consistent oscillatory mechanism for all cells. It is therefore plausible to assume that the timekeeper should be internal to the cell, whereby such an oscillator would also have the capacity to adapt to various environmental cycles through feedback loops (Mondragon-Palomino et al., 2011).

What timescale should an internal pacemaker be able to generate? Conserved biological functions range from the picosecond ($10^{-12}$ s) to the second/minute scale (**Figure 2**). These include picosecond formation of hydrogen bonds, picosecond energy transfer across phospholipid membrane leaflets (Mashaghi et al., 2012), microsecond ($10^{-6}$ s) folding of a protein (Chung et al., 2012), subsecond ($10^{-1}$ s) translation (per amino acid) in mouse embryonic stem cells (Ingolia et al., 2011) and tens of seconds ($10^{2}$ s) of search time by the yeast transcription initiation complex (Larson et al., 2011). From the time-paths presented in **Figure 2**, the shortest timescale at which chemistry is bridged to a biological task in the cell appears to be the microsecond fold time of a single amino acid. A cellular oscillator should therefore operate at approximately this frequency or higher.

Metabolic cycles, through the generation of redox-capable by-products (Wang et al., 2012), and cellular energy cycles can be proposed as sources of induced periodicity. However, they are highly variable across species and, for example, the rate-limiting step of glycolysis, the transfer of a phosphate group from phosphoenol pyruvate (PEP) by pyruvate kinase (Harris et al., 2012), would be too slow. Similarly, the cell cycle, in which the protein phosphatase Cdc14 acts as a main switch (Wittenberg, 2012), operates on much longer timeframes. On a genome level, transcriptional bursting is very gene-specific (Suter et al., 2011), and is initially stochastic followed by a steady elongation (Nair and Raj, 2011). It therefore cannot provide a species-conserved and steady pulse.

The remaining alternative is the plasma membrane. Inherent water-dependent picosecond vibrations in the phospholipid membrane (Mashaghi et al., 2012) are well shorter than the reported protein fold-time, and the membrane's position at an interfacial juncture allows it to couple the 'outside' to the 'inside' environment. This would be an ideal characteristic for a theoretical timekeeper. The membrane as a pacemaker, through contact, can also fulfill the essential task of synchronization between cells, either in unicellular or multicellular organisms. This could explain observations that cell density can affect cell-to-cell variability (Snijder et al., 2009).

This hypothesis can be refined if two pieces of evidence become available: First, how consistent is the frequency of oscillation of the plasma membrane over seconds? Furthermore, is the frequency conserved amongst cells of different phyla considering the different composition of phospholipid bilayers? What role, if any, do key phospholipids, such as phosphatidylinositol 4-phosphate (Hammond et al., 2012), play in generating this oscillation? If these and similar questions confirm the timekeeping potential of the membrane, it would be exciting to propose experiments where carefully-controlled micro-pulsation of a cell culture medium, for example, is followed by telomere length comparisons, or developments in synthetic pulse generators (Khalil and Collins, 2010) are used to reconstruct cellular timekeeping *in vitro*. Finally, utilization of advances in modeling theory may also be fruitful, whereby the oscillations of an as-yet-unformed lipid membrane is modeled in a 'chaotic' state, which then synchronize into periodicity once the cell's full 'sphere' is formed. Similar models have previously been proposed for the 'not-aroused' state of the brain (Pfaff and Banavar, 2007), which emerges out of a chaotic phase when it is aroused. Overall, understanding time at the cellular level can, for example, contribute a new layer of analysis in cancer or neurodegenerative disease research.

## ACKNOWLEDGEMENTS
I would like to thank Hezhen Ren for help with the illustrations.
## ACKNOWLEDGEMENTS
I would like to thank Hezhen Ren for help with the illustrations.

## FIGURE LEGENDS

**Figure 1. Possible external forces acting on a cell.** Light/dark cycles, sound, gravity and/or magnetism, temperature or cell-cell contact forces cannot be suggested as providing a consistent and species-conserved source of oscillation as a timekeeper for biological functions.

**Figure 2. Timescale of select cellular functions.** A logarithmic depiction of essential chemical and biological cellular tasks, based on published results, demonstrates a picosecond to second time range. Energy transfer across phospholipid membrane leaflets appears to be a common denominator of slower biological functions.

**Figure 1**

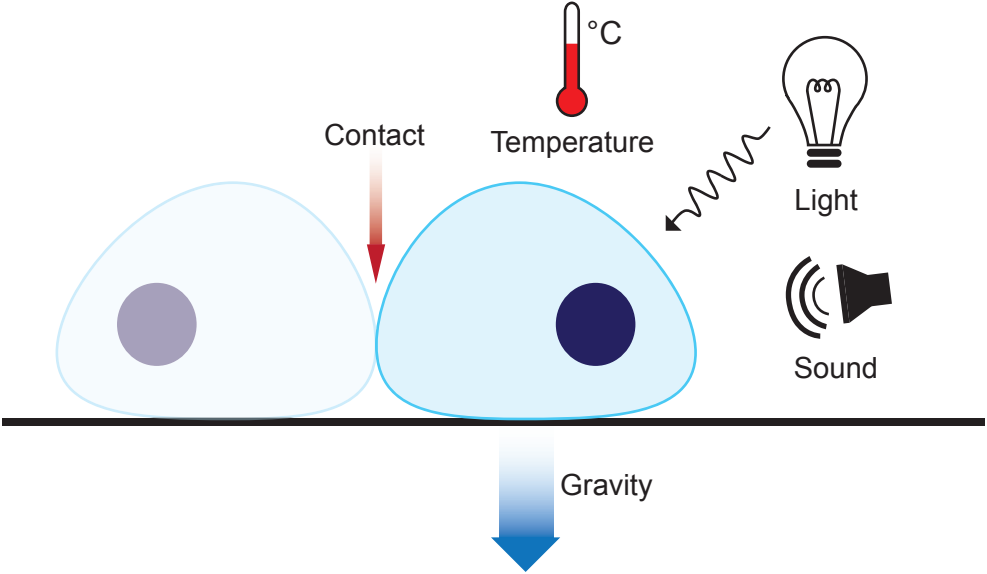

# Figure 2

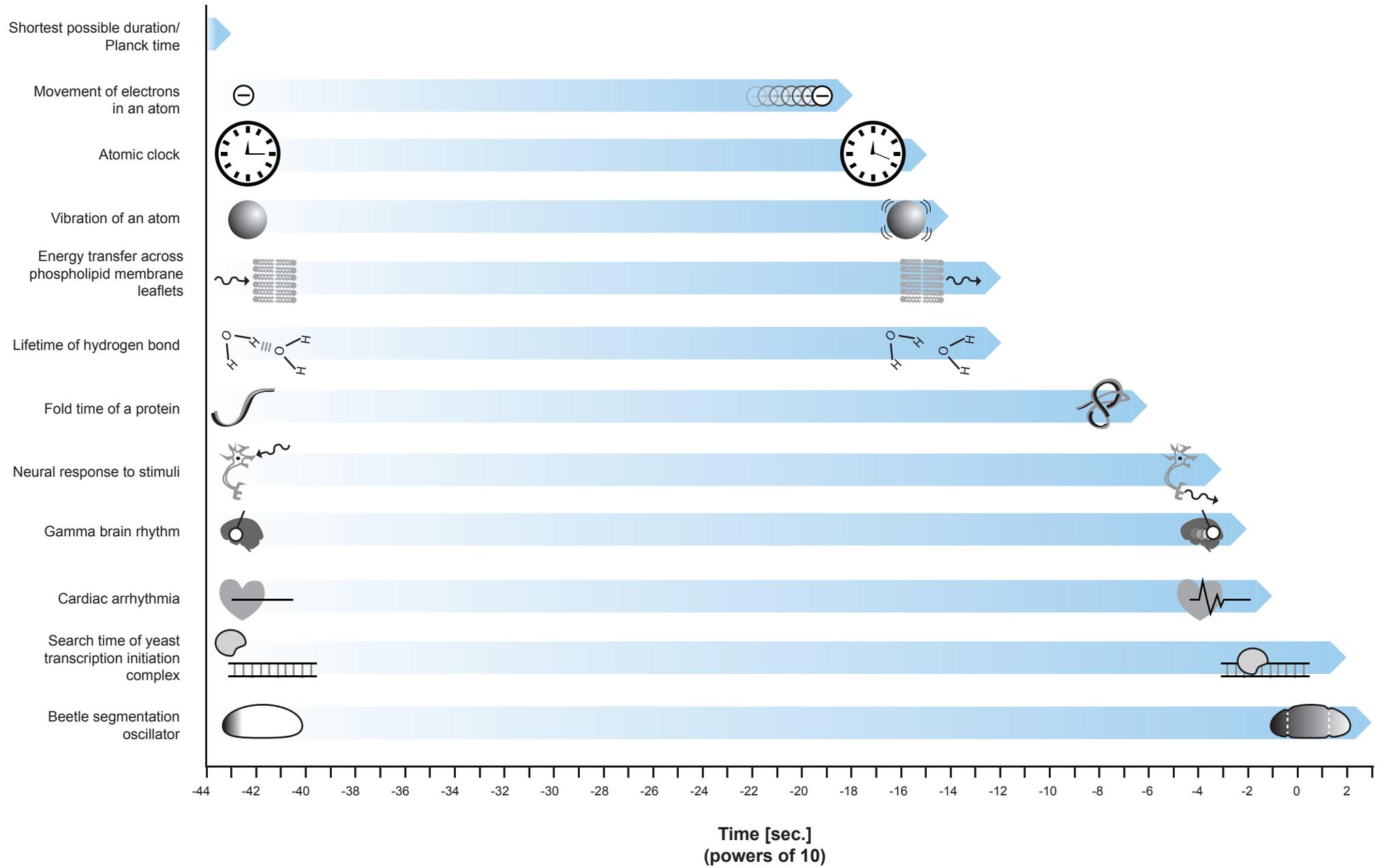